\documentclass[10pt,conference]{IEEEtran}
\IEEEoverridecommandlockouts
\usepackage[utf8]{inputenc}
\usepackage{cite}
\usepackage{verbatim}
\usepackage{amsmath}
\usepackage{amssymb}
\usepackage[dvipsnames]{xcolor}
\usepackage{amsthm}
\usepackage{booktabs}
\usepackage{mathtools}
\usepackage{slashed}
\usepackage{amsfonts}
\usepackage{makecell}
\usepackage{tabularx}
\usepackage{graphicx}
\usepackage{pgfplots}
\pgfplotsset{compat=1.18}
\usepackage{comment}
\usepackage{subfigure}
\usepackage{float}
\usepackage{color}
\usepackage{xcolor}
\usepackage{multirow}
\usepackage{physics}
\usepackage{quantikz}
\usepackage{algpseudocode}
\usepackage{algorithm}
\usepackage{tikz}
\usepackage{url}
\usetikzlibrary{shapes, arrows.meta, positioning}
\usepackage{rotating}
\def\BibTeX{{\rm B\kern-.05em{\sc i\kern-.025em b}\kern-.08em
    T\kern-.1667em\lower.7ex\hbox{E}\kern-.125emX}}
\begin{document}

\title{Quantum Computing Applications for Flight Trajectory Optimization}

\author{\IEEEauthorblockN{Henry Makhanov\IEEEauthorrefmark{1}\IEEEauthorrefmark{2}\IEEEauthorrefmark{9}, Kanav Setia\IEEEauthorrefmark{1}, Junyu Liu\IEEEauthorrefmark{1}\IEEEauthorrefmark{3}\IEEEauthorrefmark{4}\IEEEauthorrefmark{5}\IEEEauthorrefmark{6}, Vanesa Gomez-Gonzalez\IEEEauthorrefmark{7}, Guillermo Jenaro-Rabadan\IEEEauthorrefmark{7}}
\IEEEauthorblockA{\IEEEauthorrefmark{1}qBraid Co., Chicago, IL 60615, USA}
\IEEEauthorblockA{\IEEEauthorrefmark{2}Department of Computer Science, The University of Texas at Austin, Austin, TX 78712, USA}
\IEEEauthorblockA{\IEEEauthorrefmark{3}Pritzker School of Molecular Engineering, The University of Chicago, Chicago, IL 60637, USA}
\IEEEauthorblockA{\IEEEauthorrefmark{4}Department of Computer Science, The University of Chicago, Chicago, IL 60637, USA}
\IEEEauthorblockA{\IEEEauthorrefmark{5}Chicago Quantum Exchange, Chicago, IL 60637, USA}
\IEEEauthorblockA{\IEEEauthorrefmark{6}Kadanoff Center for Theoretical Physics, The University of Chicago, Chicago, IL 60637, USA}
\IEEEauthorblockA{\IEEEauthorrefmark{7}Acubed by Airbus, 601 W California Ave, Sunnyvale, CA 94086}
E-mail: \IEEEauthorrefmark{9}makhanov@utexas.edu}
\maketitle
\begingroup
\renewcommand\thefootnote{}
\footnotetext{\copyright~2024 IEEE. Personal use of this material is permitted. 
DOI: 10.1109/QCNC62729.2024.00019.}
\endgroup
\begin{abstract}
Major players in the global aerospace industry are shifting their focus toward achieving net carbon-neutral operations by 2050. A considerable portion of the overall carbon emission reduction is expected to come from new aircraft technologies, such as flight path optimization.  In pursuing these sustainability objectives, we delve into the capacity of quantum computing to tackle computational challenges associated with flight path optimization, an essential operation within the aerospace engineering domain with important ecological and economic considerations. In recent years, the quantum computing field has made significant strides, paving the way for improved performance over classical algorithms. In order to effectively apply quantum algorithms in real-world scenarios, it is crucial to thoroughly examine and tackle the intrinsic overheads and constraints that exist in the present implementations of these algorithms. Our study delves into the application of quantum computers in flight path optimization problems and introduces a customizable modular framework designed to accommodate specific simulation requirements. We examine the running time of a hybrid quantum-classical algorithm across various quantum architectures and their simulations on CPUs and GPUs. A temporal comparison between the conventional classical algorithm and its quantum-improved counterpart indicates that achieving the theoretical speedup in practice may necessitate further innovation. We present our results from running the quantum algorithms on IBM hardware and discuss potential approaches to accelerate the incorporation of quantum algorithms within the problem domain.
\end{abstract} 

\section{Introduction}\label{sec:intro}

The aviation industry plays a critical role in the global economy by enabling the movement of people and goods across countries and continents. In 2019, commercial airlines transported over 4.5 billion passengers worldwide, showcasing the industry's significance in connecting businesses, promoting tourism, and facilitating cultural exchange \cite{WorldBank2021}. With the rapid expansion of global air travel, the aviation industry has become a vital component of modern life, offering unprecedented mobility and accessibility to people worldwide.

However, this expansion comes with a considerable environmental cost. In the European Union alone, the aviation industry is responsible for approximately 3.8\% of the anthropogenic $\text{CO}_2$ emissions \cite{EuropeanCommission2021}, and its fuel consumption is projected to increase four to six times between 2010 and 2050 \cite{ICAO2019}. Furthermore, air travel produces other greenhouse gases, such as nitrogen oxides and water vapor, contributing to climate change. The environmental impact of commercial aviation has become a pressing ecological issue, and addressing the industry's carbon footprint is crucial to mitigating climate change and achieving global sustainability goals. Thus, global aviation faces highly complex technical problems with important environmental and economic consequences. 

In this work, we will explore a quantum-enhanced algorithm to find the most fuel-efficient path between source and destination, thus solving the flight path optimization problem. Optimized flight paths can not only help minimize greenhouse gas emissions, but they can also effectively alleviate the environmental impact of the increasing air traffic demand, which is predicted to rise in the coming decades \cite{ICAO2019}. Moreover, enhanced flight path optimization has the potential to reduce contrail formation, which is another contributor to climate change \cite{burkhardt2008contrail}.  We have assumed that the fleet assignment problem has already been solved, an optimal aircraft has been assigned, and a landing spot will be available upon the aircraft's arrival at the destination airport \cite{cacchiani2013heuristic}. Flight path optimization remains a complex problem  with significant challenges:

\begin{itemize}
\item Complexity and High Dimensionality: Flight optimization is a multi-variable optimization problem with numerous constraints, such as fuel consumption, flight time, aircraft weight, and air traffic control restrictions. The intricacy of the problem stems from the multitude of interconnected variables, which contribute to its high dimensionality. This aspect poses a challenge for traditional optimization techniques, as the vast solution space makes finding the optimal solution a time-consuming task.
\item Dynamic Environment and Real-time Constraints: Aircraft operate under rapidly changing conditions, such as fluctuations in the wind, air traffic control restrictions, and equipment failures, which create a dynamic and uncertain environment \cite{cook2011european}. These complexities necessitate the development of accurate models and algorithms that can adapt and respond in real time. As the aircraft must decide its flight path and operating conditions while airborne, flight optimization requires high computational power and fast algorithms capable of handling large amounts of data and providing solutions within a short time frame.
\end{itemize}

Although quantum computers may not be expected to be universally useful for all computations within the next decade, several quantum algorithms already provides promising results \cite{shor1994algorithms, kitaev1995quantum, harrow2009quantum}. In our paper, our goal is to develop an architecture that integrates various computational paradigms, allowing the software to determine the optimal computing resource (CPU, GPU, or quantum) for solving each problem.

Our strategy involves modularizing flight path optimization problems and identifying instances where known quantum algorithms can provide faster solutions. However, one key challenge of these quantum algorithms is the input/output problem. Since quantum algorithms rely on properties such as quantum superposition and entanglement, they often assume the existence of a quantum database that can be queried in superposition. In other cases, these algorithms require data input in quantum superposition and return a probabilistic output. Consequently, even with a well-defined quantum algorithm, uncertainties may remain regarding data input and output interpretation. In our work, we attempt to address the engineering aspects of some of these problems to provide a more comprehensive understanding.

In this work, we investigate the potential of the minima finding variation of Grover's algorithm to replace the classical minima finding subroutine in the Dijkstra algorithm \cite{dijkstra1959note}. This algorithm is a modification of a famous Grover's unstructured search algorithm \cite{durr1996quantum}, which implies the quadratic asymptotic advantage over the classical minima finding subroutine. We addressed the weight update step by leveraging the computational tractability of the specific digraph structure of the problem and advancements in workload parallelization \cite{shi2018graph}. We benchmark the actual runtime of the quantum algorithm and offer techniques to estimate the time required to execute the complete pipeline for the problem. We then compare these results with the time needed to run the pipeline on a classical computer to evaluate the potential advantages of using quantum algorithms in flight path optimization.

In Section \ref{sec:backgrounds}, we provide an overview of different methodologies, optimal control, and non-optimal control, that are usually used to solve the flight path methods. In Section \ref{sec:architecture}, we provide details for the mathematical model based on our design choices and details for the architecture of our software. In section \ref{sec:implementation}, we talk about our methodology for solving the problem, including the details for using the Quantum Minima Finding algorithm within Dijkstra's algorithm. In Section \ref{sec:results}, we provide a summary of results and their implications. In Section \ref{sec:conc}, we conclude with a summary of our work and potential future directions.

\section{Background}\label{sec:backgrounds}

\begin{figure*}[htp]
    \centering
    \includegraphics[width=12cm]{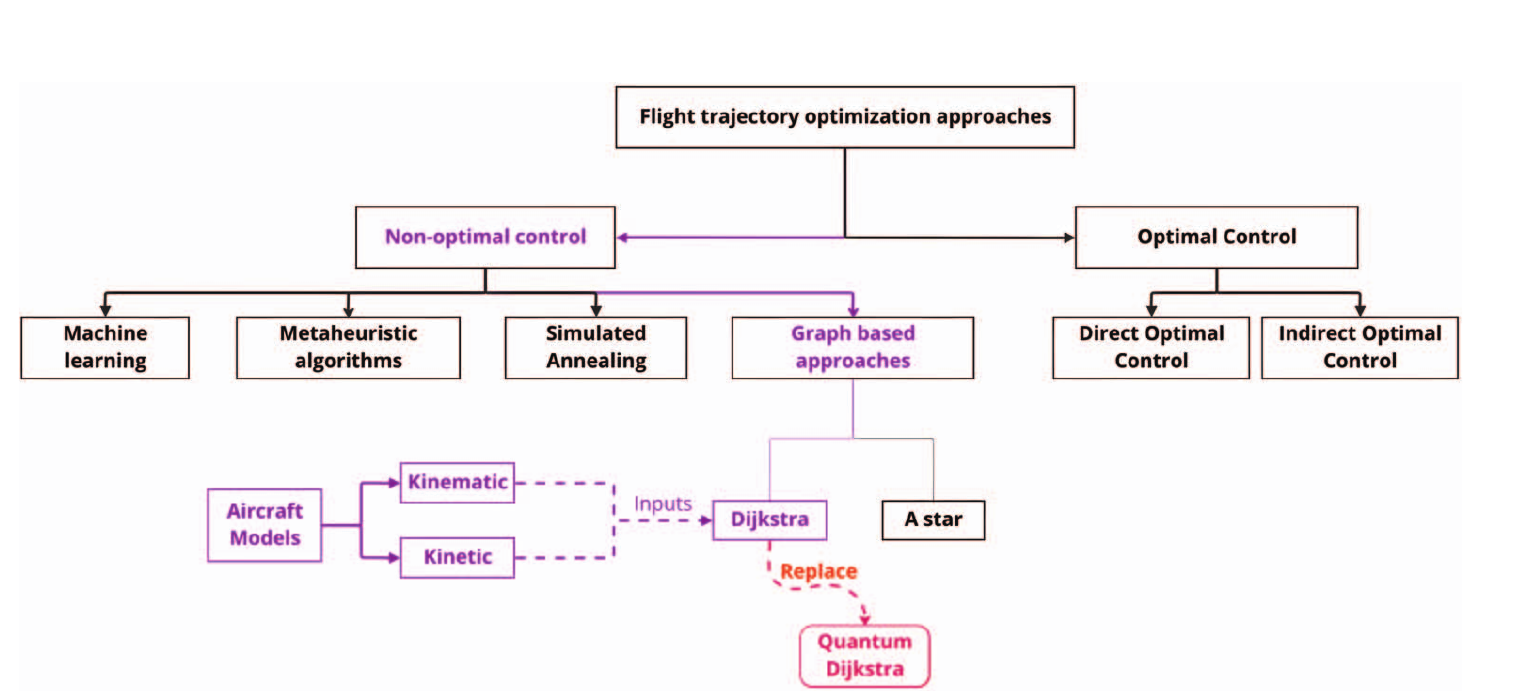}
    \caption{Diagram of methods used for flight trajectory optimization.}
    \label{fig:framework}
\end{figure*}

Flight trajectory optimization is a critical problem in the aviation industry, as it addresses several essential objectives, such as collision avoidance, fuel efficiency, and minimizing environmental impact. Given the importance of flight optimization many methodologies have been developed to address it, each tailored for specific aspects of the problem. The continuous advancement of these methodologies contributes to both the operational efficiency and environmental sustainability of the aviation sector, making flight optimization a cornerstone of modern aviation.

The taxonomy of flight optimization techniques (Figure \ref{fig:framework}) can be categorized based on the approaches used to model the problem and the mathematical representation of the aircraft. There are two primary approaches for this problem: optimal control and non-optimal control. Each approach offers its unique set of advantages and disadvantages, depending on the specific conditions and objectives of the problem being addressed.

When employing either the optimal or non-optimal control methods, it is necessary to establish a mathematical model for the aircraft. These models can be formulated using one of two fundamental approaches: kinematic or kinetic (also known as dynamic). The kinetic or dynamic approach considers all forces acting on the aircraft (such as thrust, lift, and drag) to calculate its motion through space. In contrast, the kinematic model is solely concerned with the aircraft's kinematic variables and omits any consideration of the forces involved.
\subsection{Optimal control approach}
The optimal control approach, a mathematical optimization technique, treats the aircraft as a dynamical system and focuses on minimizing a chosen objective function, such as fuel consumption, flight duration, or emissions, to solve the flight optimization problem \cite{simorgh2022comprehensive, schnitzer2015evaluation}. By identifying the optimal control parameters, including altitude, airspeed, and route, this method achieves the desired minimization while adhering to the non-linear constraints imposed by the aircraft's capabilities and airspace regulations. The problem addressed by the optimal control approach belongs to the non-linear constraint convex optimization class because it involves minimizing a convex objective function subject to non-linear constraints on the control and state variables. The convex nature of the objective function ensures that any local minimum is also a global minimum, while the non-linear constraints contribute additional intricacies to the optimization process.

To describe the aircraft's dynamics, the time evolution of its state can be modeled using differential equations that link the state of the aircraft with the control variables:
\begin{align}\label{eq:DEA}
\dot{\mathbf{x}}(t)=\mathbf{f}(t,\mathbf{p},\mathbf{x}(t),\mathbf{u}(t))
\end{align}
Here, $t$ represents time, $\mathbf{x}$ is the vector denoting the state of the system, $\mathbf{f}$ is the vector field, $\mathbf{u}$ is the vector containing all the control variables, and $\mathbf{p}$ comprises the scalar variables \cite{simorgh2022comprehensive}. Given that time is continuous and the differential equation \eqref{eq:DEA} represents the system's dynamical behavior, the optimal control method is regarded as a reliable technique for obtaining the optimal solution.

Minimizing the objective function is complicated because the state of the system differential equation \eqref{eq:DEA} necessitates solutions under non-linear constraints. Generally, these constraints are expressed as follows:
\begin{align}
\mathbf{h}(\mathbf{x}(t),\mathbf{u}(t),\mathbf{p},t)=\mathbf{0}\\
\mathbf{g}(\mathbf{x}(t),\mathbf{u}(t),\mathbf{p},t)\leq \mathbf{0}
\end{align}
Here, $\mathbf{h}$ and $\mathbf{g}$ are vector fields, $\mathbf{0}$ is a vector of zeros, and the equalities and inequalities hold in an element-wise manner \cite{simorgh2022comprehensive}. In some cases, the constraints may be non-differentiable, and there could be additional conditions on the initial and final values of the state and control variables, further complicating the optimization problem. In the context of our project, we have opted for a simpler model that would have a clearer mapping to a quantum computationally solvable problem.
\subsection{Non-optimal control} 
The non-optimal control approach comprises a diverse set of techniques distinct from optimal control methods \cite{pedersen2010tuning}. These heuristic optimization techniques include graph-based methods, nonlinear programming, and machine learning approaches \cite{van1987simulated, katoch2021review}. By relying on heuristic strategies rather than mathematical optimization, these methods offer a different approach to solving complex problems, often with reduced computational complexity and more practical implementation.

In our project, we selected Dijkstra's algorithm, a method rooted in graph theory, owing to its polynomial asymptotic complexity and notable efficiency \cite{dijkstra1959note}. This class of techniques simplifies the problem formulation, which in turn eases the implementation and analysis processes on quantum computing platforms.

The process of utilizing graph-based methods can be divided into three primary steps: (1) Graph generation, where the three-dimensional space is discretized, with the graph size and density influencing the solution's accuracy and computational complexity; (2) Edge weight calculation, which determines the cost associated with traversing the graph based on the optimization objectives (e.g., minimizing CO2 emissions, time efficiency, collision avoidance), and (3) Running the single-source shortest path (SSSP) algorithm, such as Dijkstra's algorithm, A star algorithm, or Bellman-Ford, depending on the specific problem requirements and available heuristics. 
\subsection{Kinetic and Kinematic models of aircraft performance}
Aircraft trajectory modeling, an essential aspect of flight optimization, requires the incorporation of performance models to minimize the objective function. Two main performance models exist for aircraft: the kinematic and kinetic models \cite{nuic2010bada, csanda2018using}. The kinematic model simplifies the problem by ignoring the underlying physics, while the kinetic model takes into account the various forces acting on the aircraft. 

The kinematic approach, employed when motion can be described by basic parameters like speed, direction, and altitude, is well-suited for modeling simpler trajectories \cite{sun2022openap}. Examples include the great circle route, which represents the shortest path between two points on a sphere, and the straight-line flight path, a rudimentary model used in air traffic control systems to predict an aircraft's trajectory between waypoints \cite{nastro2010great}. However, the kinematic approach does not consider external forces, such as wind or air resistance.

In contrast, the kinetic approach, which solves dynamic equations of motion to determine the aircraft's position, velocity, and acceleration as a function of time, is vital for generating more accurate and realistic predictions of aircraft behavior in real-world situations \cite{sun2020openap, sun2022openap}. While both the kinematic and kinetic models are widely used in aviation, the choice between them depends on the complexity and accuracy requirements of the specific problem.

\section{Mesh Grid Generation} \label{sec:architecture}
\subsection{Graph Construction}
\begin{figure}[h]
    \centering
    \includegraphics[width=5.5cm]{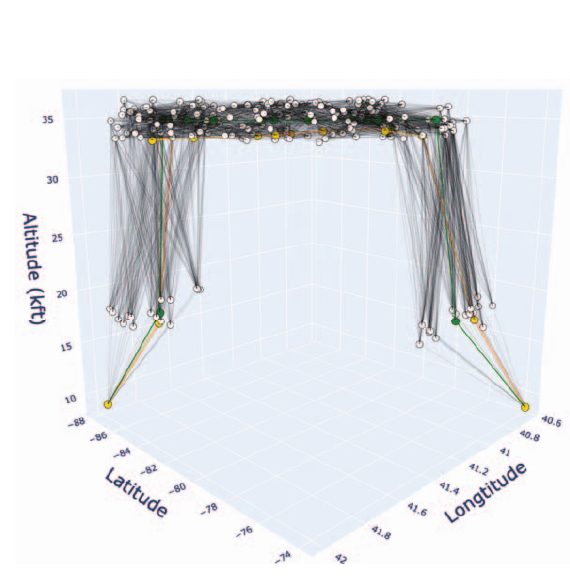}
    \caption{3D Graph of the perturbed routes from New York to Chicago, with the  single-source shortest path calculations: \textcolor{ForestGreen}{Original Path}, \textcolor{YellowOrange}{Quantum Dijkstra Path}, \textcolor{BrickRed}{Classical Dijkstra Path}. Altitude is in Kilofeet (1 kft = 1,000 feet).}
    \label{fig:3Dgraph}
\end{figure}
The full graph generation process constructs a comprehensive network of flight routes between origin and destination airports using a Directed Acyclic Graph to represent the mesh grid of interconnected paths. We constructed the mesh grid with a three-step process: 

\begin{enumerate}
    \item Application Programming Interface call is made to the Flight Plan Database to acquire the flight route, consisting of the latitude, longitude, and maximum altitude between the specified airports \cite{flightplan_database}.
    \item The original route is then perturbed and transposed into three dimensions using the altitude profile function. The granularity of perturbation, specified for latitude, longitude, and altitude, allows for controlled modifications of the route. By iterating through a predetermined number of cycles, multiple perturbed paths are generated, each representing a potential variation of the original route.
    \item Directed Acyclic Graph is constructed, which includes both the original and perturbed paths. The graph is created using the NetworkX library \cite{networkx}. The nodes and edges are added to the graph, maintaining the connections between the original nodes and the perturbed nodes in the subsequent layers.
\end{enumerate}

The mesh grid is visualized in a 3D plot using specialized libraries: Plotly, NetworkX, and NumPy (Figure \ref{fig:3Dgraph}) \cite{plotly, networkx, numpy}. A 3D spring layout of the mesh grid is generated, where node positions are replaced with actual coordinates from the paths. Separate traces are created for nodes and edges of the directed graph, as well as distinct traces for the original path, Dijkstra's path, and A star path. Each path is represented by separate node and edge traces with unique colors for easy differentiation. Wind information can be incorporated into the visualization using vector representations. This visualization approach provides a clear and interactive representation of the mesh grid and its associated paths, allowing for a deeper understanding and analysis of the underlying structure.
\subsection{Calculating the weights of the graph}
The selection of edge weight functions is contingent upon the optimization objectives, which could include factors such as time, fuel consumption, and carbon dioxide emissions. In the present study, we primarily aimed to minimize fuel consumption throughout the entire flight trajectory. Since fuel consumption is strongly correlated with carbon dioxide emissions, this optimization indirectly contributes to reducing emissions as well \cite{EPAOverviewGHG}.

Each aircraft possesses a distinct performance model that defines its fuel consumption function. For this research, the Airbus A320 was selected, a very popular and reliable aircraft, with over 5,000 units currently in operation \cite{AirbusOrdersDeliveries}. The CFM International CFM56-5B4
the engine was chosen to accompany this aircraft \cite{CFM56EngineFamily}. Table \ref{tab:aircraft_data} presents all the relevant parameters for the aircraft, which are crucial in determining its fuel consumption and in the computation of Thrust Specific Fuel Consumption (TSFC). The fuel consumption model and other empirical factors required for calculation were taken from \cite{campbell2010multi, sun2022openap}.

\begin{table}[ht]
    \centering
    \caption{Parameters used for the optimal flight trajectory model Airbus A320 with CFM56-5B4 engine.}
    \begin{tabular}{lcc}
        \hline
        \textbf{Parameter} & \textbf{Value} & \textbf{Units} \\
        \hline
        Maximum Takeoff Weight (MTOW) & 77,000 & kg \\
        Maximum Landing Weight (MLW) & 64,500 & kg \\
        True Airspeed (TAS) & 500 & mph \\
        Drag Coefficient (CD) & 0.022 & - \\
        Reference Area (S) & 122.6 & m$^2$ \\
        \hline
    \end{tabular}
    \label{tab:aircraft_data}
\end{table}

When the aircraft is at cruise altitude and velocity, the fuel consumption ($W_e$) between two nodes can be calculated using the following equation:
\begin{align*}
    W_e = \frac{W_f\times d_e}{V_{e}} 
\end{align*}
Here, $W_f$ represents fuel flow, $d_e$ denotes the distance between two nodes based on the Euclidean metric, and $V_{e}$ is the average velocity. The fuel flow at cruise altitude and velocity is expressed as:
\begin{align*}
    W_f = T_{\text{req}} \times  \text{TSFC}~
\end{align*}
where $T_{\text{req}}$ is the thrust required, and TSFC is thrust-specific fuel consumption. The thrust required in cruise mode is the sum of the clean drag ($D$) on the aircraft and the component of the weight of the aircraft in the direction opposite of thrust: 
\begin{align}
T_{req} = D + m \times g \times \sin(\gamma)~
\end{align}
where, $m$ is the mass of the aircraft, $g$ is the gravitational constant and $\gamma$ is the flight path angle in radians. 

Fuel consumption during the ascent and descent phases of the flight can deviate significantly from the aforementioned expressions, as drag depends on the aircraft's velocity, polarity, and air density. The drag polar is also a nonlinear function specific to a particular aircraft. The fuel flow during ascent and descent is expressed as:
\begin{align}
W_f = W_{\text{sea-level}} + W_{\text{altitude-adjusted}}~,
\end{align}
where, $W_\text{sea-level}$ is the fuel flow at the sea level, and $W_{\text{altitude-adjusted}}$ is the fuel flow adjusted from the altitude. 
Both $W_\text{sea-level}$ and $W_{\text{altitude-adjusted}}$ are 3rd-degree polynomials whose coefficients could be obtained empirically. In our work, we focus on the cruise phase of the flight and assume constant velocity throughout the flight.

\begin{figure*}[htp]
    \centering
    \includegraphics[width=9.5cm]{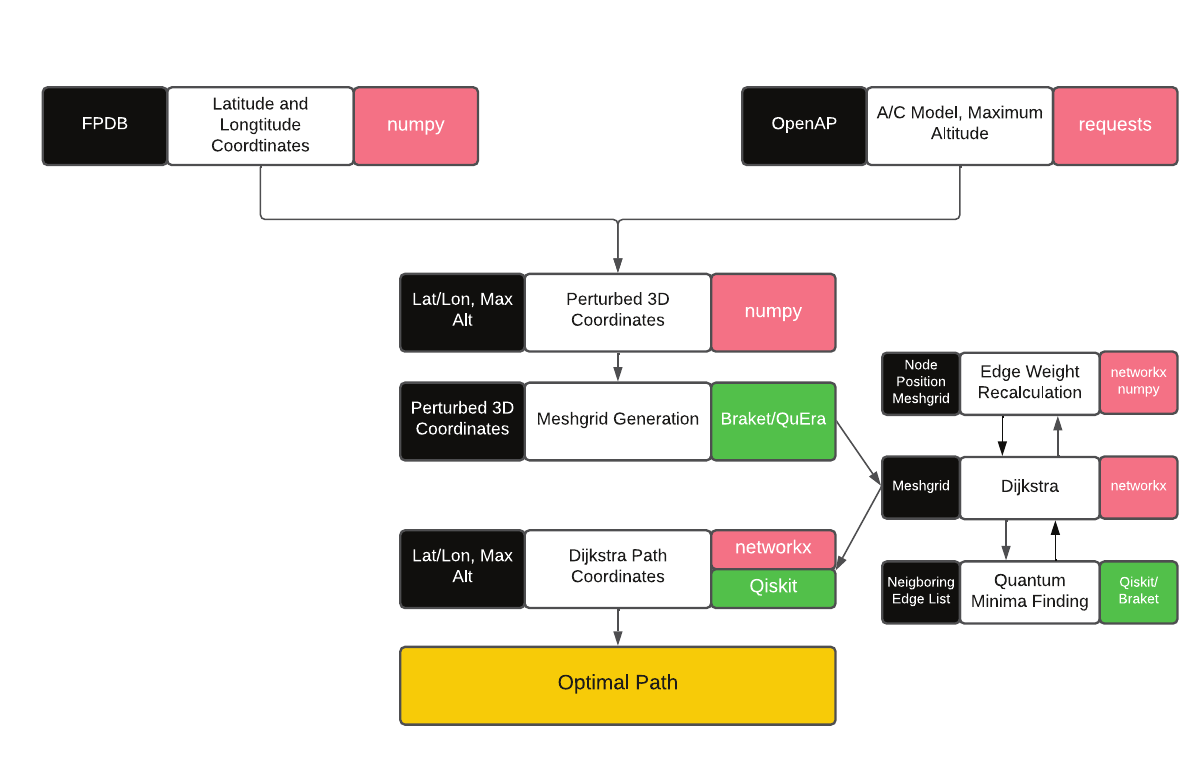}
    \caption{Graphical depiction of the software implementation of Quantum-enhanced Dijkstra.}
    \label{fig:software}
\end{figure*}

\section{Single-source Shortest Path }\label{sec:implementation}
In the previous section, we explained the mechanisms of mesh-grid construction, directed graph representation, and edge weight recalculation in the context of our problem. We now will explore Dijkstra's algorithm and it's quantum-enhanced version for solving the single-source shortest path problem.
\subsection{Classical Dijkstra} \label{ss:dijkstra}
Dijkstra's algorithm was developed by Edsger Dijkstra in 1956 and since became a fundamental tool in graph analysis \cite{dijkstra1959note}. Given a graph $G=(V,E)$, where $V$ is the set of vertices and $E$ is the set of edges with an associated edge weight function $w$, Dijkstra's algorithm finds the shortest path from a source vertex $s$ to a destination vertex $d$, with both $s$ and $d$ being elements of the vertex set $V$.

The algorithm works by iteratively exploring neighboring vertices, updating the shortest path and distance to each vertex. The process terminates when it has found the shortest path to the destination node or has determined that no path between the two nodes exists. The algorithm starts by initializing the distance estimates for all nodes to be infinity, except for the source node, which is set to zero. The algorithm then repeatedly extracts the node with the lowest priority from the priority queue, updates the distance estimates of its neighbors based on the distance to the current node, and adds these neighbors to the priority queue.

The original implementation of Dijkstra's algorithm used an array to implement the priority queue, resulting in a time complexity of $\mathcal{O}(V^2)$ \cite{Schrijver2012}. Most efficient implementation of Dijkstra's algorithm uses the Fibonacci heap priority queue, which has a time complexity of $\mathcal{O}(E + V\log V)$ \cite{FredmanTarjan1987}. Currently, there are no quantum algorithms that can beat the Fibonacci heap Dijkstra in terms of time complexity \cite{krauss2020solving}. However, there are cases where using an unsorted array can be more practical than using more complex data structures, especially for problem sizes with low number of intermediate layers, such as short-distance, low-altitude routes. In such cases, the ease of implementation makes the unsorted array a more suitable candidate for solving the problem.
\begin{center}
\begin{algorithm}[H]
\caption{Dijkstra's algorithm for weighted DAGs}\label{alg:dijkstra-dag}
\begin{algorithmic}[1]
\Function{Dijkstra-DAG}{$G=(V,E), s$}
\State $\text{dist}(s) \gets 0$, $S \gets \emptyset$, $Q \gets \text{empty priority queue}$
\For{$v \in V$}
\If{$v \neq s$}
\State $\text{dist}(v) \gets \infty$
\EndIf
\State insert $v$ into $Q$ with key $\text{dist}(v)$
\EndFor
\While{$Q \neq \emptyset$}
\State $u \gets$ node in $Q$ with minimum key
\State move $u$ from $Q$ to $S$.
\For{$(u,v) \in E$}
\State $\text{alt} \gets \text{dist}(u) + w(u,v)$
\If{$\text{alt} < \text{dist}(v)$}
\State $\text{dist}(v) \gets \text{alt}$
\State update key of $v$ in $Q$ to $\text{dist}(v)$
\EndIf
\EndFor
\EndWhile
\State \Return $\text{dist}$
\EndFunction
\end{algorithmic}
\end{algorithm}
\end{center}
\subsection{Quantum-enhanced Dijkstra} \label{ss:quantum-dijkstra}

This project focuses on improving the theoretical performance of the classical Dijkstra algorithm. In Figure \ref{fig:software}, we outline the full software architecture of our project. Dijkstra algorithm contains two subroutines that contribute to its $\mathcal{O}(V^2)$ complexity: Minima finding on the unstructured list, and weights update in the priority queue $Q$, lines 10 and 12-17, in Algorithm 1, respectively. We address the minima finding subroutine with the minima finding a variation of Grover's quantum search algorithm for the minima finding \cite{durr1996quantum}. For the weights update in the priority queue, we used parallelism to achieve a simultaneous update at a constant computational cost.
\subsubsection{Minima finding variation of Grover's algorithm} \label{ss:Grover}
Grover's algorithm, known for its quadratic speedup in searching unsorted databases compared to classical algorithms, is adapted in our work to find the minimum value in an unsorted list. Consider an array of $N$ elements, where $N=2^n$ for simplicity. Our algorithm is an implementation of the Dürr-Høyer minima finding algorithm \cite{durr1996quantum}. We define a binary function $f: \{0, 1, \ldots, N-1\} \to \{0, 1\}$ such that $f(i) = 1$ if the $i^{th}$ element of $array$ is less than the current minimum guess, and $f(i) = 0$ otherwise.

Algorithm has two distinct Minimum Oracles: the Minimum Oracle for Array ($U_{array}$) and the Minimum Oracle for Initial Guess ($U_{guess}$). The $U_{array}$ is designed to interact with the quantum states corresponding to the array elements, marking those states where the elements are less than a given threshold, which is chosen arbitrarily. On the other hand, $U_{guess}$ is utilized to establish an initial threshold or reference point against which the array elements are compared.

\begin{center}
\begin{algorithm}
\caption{Quantum Minimum Finding based on Grover's Algorithm}
\label{quantum_minimum_finding_compact}
\begin{algorithmic}[1]
\Procedure{QuantumMinimumFinding}{$array$}
\State $N \gets \text{length}(array)$
\State $\text{num qubits} \gets \lceil \log_2 (N) \rceil + 1$
\State $\text{initial guess} \gets \text{RandomInteger}(0, N-1)$
\State $\text{done} \gets \text{True}$
\While{$\text{done}$}
\State Initialize superposition $\ket{s} \gets \frac{1}{\sqrt{N}} \sum_{i=0}^{N-1} \ket{i}$
\State Define Minimum Oracle for $array$
\State Apply Minimum Oracle to $\ket{s}$ resulting in $\ket{s'}$
\State $\text{iterations} \gets \left\lfloor \frac{\pi}{4}\sqrt{\frac{N}{\text{num marked elements}}} \right\rfloor$
\For{$j \gets 1$ to $\text{iterations}$}
\State Oracle: $\ket{i} \gets (-1)^{f(i)} \ket{i}$
\State Diffusion: $\ket{s'} \gets 2 \ket{s'} - \frac{2}{N} \sum_{i=0}^{N-1} \ket{i}$
\EndFor
\State Measure $\ket{s'}$ to obtain the marked state
\State $\text{next guess} \gets \text{MostCommonOutcome}(\ket{s'})$
\If{$array[\text{next guess}] < array[\text{initial guess}]$}
\State $\text{initial guess} \gets \text{next guess}$
\Else
\State $\text{done} \gets \text{False}$
\EndIf
\EndWhile
\State \Return $array[\text{initial guess}]$
\EndProcedure
\end{algorithmic}
\end{algorithm}
\end{center}

The algorithm starts with the establishment of an equal superposition state $\ket{s} = \frac{1}{\sqrt{N}} \sum_{i=0}^{N-1} \ket{i}$, which represents an equal probability of all possible indices of the array. To transform this state into $\ket{s'}$, the $U_{array}$ is applied to $\ket{s}$. This oracle marks the elements less than the current threshold, altering the superposition state to $\ket{s'}$, where the marked states are phase-inverted compared to the unmarked states. The state $\ket{s'}$ thus contains the quantum representation of the array elements that are candidates for being the minimum, based on the current threshold set by $U_{guess}$ if it has been used. Grover's algorithm is applied iteratively, involving $U_{min}$ and the diffusion operator, to amplify the probability of observing the desired marked state. Repeated measurements are made to statistically identify the most common outcome, ensuring a high probability of finding the minimum index. The overall complexity of this process is $O(\sqrt{N})$, leveraging the quantum advantage for efficient minimum value search in unsorted arrays \cite{durr1996quantum, Durr2006}. The algorithm's complexity is $O(\sqrt{N})$, which offers a quadratic speedup compared to classical search algorithms for finding the minimum element in an unsorted list. The optimal number of iterations for probability amplification is approximately $\left\lfloor \frac{\pi}{4}\sqrt{\frac{N}{M}} \right\rfloor$, where $M$ is the number of marked elements. After completing the iterations, a measurement is performed on the final state, yielding the index of the new minimum guess with high probability. The algorithm proceeds iteratively, updating the current minimum guess until convergence. This adaptation of Grover's algorithm effectively exploits the quadratic speedup to enhance the efficiency of the minima-finding process.

\subsubsection{Parallel Weight Update}
The classical Dijkstra algorithm's $\mathcal{O}(V^2)$ complexity arises from two components: searching for the minimum value in an unstructured list, which takes $\mathcal{O}(V)$ time, and updating weights of adjacent nodes, which also takes $\mathcal{O}(V)$. However, by exploiting the properties of the directed acyclic graph (DAG) structure, we can improve the algorithm's performance.

Consider a DAG $G = (V,E)$ with $n$ nodes arranged in $k$ layers, where nodes are only connected to their immediate adjacent nodes in the next and previous layers. Let $|V_i|$ denote the number of nodes in layer $i$. To calculate the density of $G$, we first determine the number of edges in $G$, considering edges between adjacent layers and edges within each layer. The density of $G$ is then given by $\operatorname{density}(G)=\frac{|E|}{\text { max edges }}$. Since each node is only connected to its immediate adjacent nodes in the next and previous layers, the density of $G$ is smaller than 1, which can be exploited for faster algorithm performance \cite{ahuja1993network}.

The weight update step in Dijkstra's algorithm can be parallelized due to the sparse nature of the graph, allowing constant time per node. For each node $u \in V$, we need to update the weights of its adjacent nodes $v \in V$ such that the shortest path from the source node to $v$ traverses $u$. The relatively small number of adjacent nodes for each node $u$ enables efficient parallelization of the weight update step \cite{kirk2010programming}.

By allocating each node $u$ to a distinct processing unit (e.g., a CPU core or GPU thread), we can concurrently update the weights of its adjacent nodes $v$. Linear algebra operations, such as matrix-vector multiplication or sparse matrix multiplication, can be used to achieve this task, as they are well-suited for efficient parallelization on modern GPUs \cite{bell2008efficient}.

Contemporary GPUs are designed to execute numerous parallelizable tasks simultaneously, making them ideal for computations on large, sparse graphs. For example, the NVIDIA Tesla V100 GPU has over 5,000 CUDA cores and can perform 7.8 teraflops of floating-point operations per second \cite{nvidia_v100}. This degree of parallelism enables efficient multithreading, ensuring the weight update step's completion in constant time per node, even for large graphs \cite{kirk2010programming}.
\subsubsection{Asymptotic Analysis}
We can put our techniques together to present a modified version of Dijkstra's algorithm that uses a quantum computing-assisted minima finding function and leverages GPU multithreading to parallelize weight recalculation. Our modifications are highlighted in green, in the Algorithm 3. Recall that in the original Dijkstra, the time complexity of $\mathcal{O}(V^2)$ arises from the minima finding step and the update step.

\begin{center}
\begin{algorithm}
\caption{Quantum-enhanced Dijkstra's algorithm for weighted DAGs, with parallelized update weight update subroutine.}\label{alg:quantum-dijkstra-dag}
\begin{algorithmic}[1]
\Function{Dijkstra-DAG}{$G=(V,E), s$}
\State $\text{dist}(s) \gets 0$
\State $S \gets \emptyset$
\State $Q \gets \text{empty priority queue}$
\For{$v \in V$}
\If{$v \neq s$}
\State $\text{dist}(v) \gets \infty$
\EndIf
\State insert $v$ into $Q$ with key $\text{dist}(v)$
\EndFor
\While{$Q \neq \emptyset$}
\State \textcolor{ForestGreen}{use QMF to find minimum node $u$ in $Q$.}
\State \textcolor{ForestGreen}{move $u$ from $Q$ to $S$.}
\For{$(u,v) \in E$ \textcolor{ForestGreen}{\textbf{in parallel}}}
\State $\text{alt} \gets \text{dist}(u) + w(u,v)$
\If{$\text{alt} < \text{dist}(v)$}
\State $\text{dist}(v) \gets \text{alt}$
\State \textcolor{ForestGreen}{update key of $v$ in $Q$}
\EndIf
\EndFor
\EndWhile
\State \Return $\text{dist}$
\EndFunction
\end{algorithmic}
\end{algorithm}
\end{center}

In our algorithm, the update step vanishes because its constant complexity is dominated by the minima finding terms. Leaving us with time complexity $\mathcal{O}(\sqrt{C} \times L)$, where $C$ is the cardinality of the graph(number of nodes in the intermediate layers), and $L$ is the number of nodes in the historical graph, used for meshgrid generation (which also corresponds to the number of layers in the final graph). For our graph, the total amount of vertices is $V = C \times (L - 2) + 2$. Thus we have an improvement of $\sqrt{C}$, when compared to the original Dijkstra's algorithm, which uses unsorted lists as its priority queue.

\section{Results}\label{sec:results}
The goals for our simulations were as follows:
\begin{itemize}
    \item Validate Quantum Dijkstra algorithm with classical Dijkstra using simulators along with understanding the performance of simulators on CPU and GPU.
    \item Compare the running time of the Quantum Minima Finding algorithm on different quantum architectures. Discover if there's a quantum architecture that emerges as a clear winner.
\end{itemize}

The subsequent subsections provide our results for the quantum minima finding algorithm on both the simulator and real hardware to assess its potential as a subroutine in Dijkstra's algorithm. We have counted the total quantum gates in circuits for running quantum minima finding on the lists of various lengths, thus allowing us to calculate the theoretical feasibility of the quantum minima finding algorithm, for 3 different qubit modalities, without the operational noise associated with the background software implementations.
\subsection{Simulator Performance Analysis}
We simulated the Quantum Minima Finding algorithm for finding the minimum element of a list on different list sizes, represented as $2^N$, where $N$ denotes the number of qubits ranging from 2 to 8 (i.e., 4 to 256). We compared the performance of CPU and GPU (with NVIDIA CuQuantum) simulations using density matrix simulators to gain insights into their relative efficiency under varying conditions \cite{nvidia_cuquantum}. We used the density matrix simulator because it can handle mixed quantum states and account for various noise sources, making them more suitable for simulation of the real-world quantum computing applications. Our findings indicate that while the
\begin{figure}[H]
    \centering
    \includegraphics[width=8.5cm]{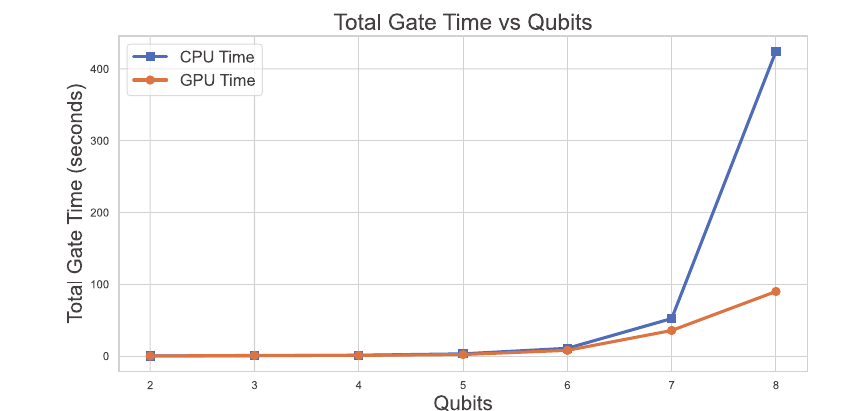}
    \caption{GPU vs CPU total gate time performance on list lengths of 4 to 256 ($2^2$ to $2^8$)}
    \label{fig:cpu_gpu_gate_time}
\end{figure}
 CPU and GPU simulations have comparable performance for smaller list sizes, the differences become more pronounced as the list size increases, particularly when the list length reaches 128 ($2^7$) or larger. We have visualized our findings in Figure \ref{fig:cpu_gpu_gate_time}. Despite that, given the operational degree of the graph, list lengths of more than 100 are unrealistic. Therefore, we maintain that the CPUs provide a competitive level of performance in comparison with the GPU for the majority of realistic flight simulation experiments.
\subsubsection{The Gate speed performance vs. Classical Minima finding}
We recognized that the performance of the algorithm is affected by the operational overhead, such as library imports. Additionally, the native Python minima function has an enhanced performance due to its low-level C language implementation. We were interested in seeing the theoretical limit of how well the Quantum Minima Finding algorithm could perform in ideal conditions. We counted the total number of gates it takes to run a Quantum Minima Finding algorithm and transpiled it into one or two-qubit set $CNOT$, $SWAP$, $H$, $|+\rangle$ state preparation, $|0\rangle$ state preparation, $X$ measurement, $Z$ measurement, $X, Y, Z, S, T$. We then multiplied the corresponding gate counts by the known gate execution times of 3 qubit modalities (Superconducting, Trapped Ion, and Neutral Atoms) \cite{suchara2013comparing}. This approach, however, needs to be taken as an upper bound, as in practice, oftentimes gates can operate in parallel, offering faster times then calcuated using our method. Our investigation revealed that superconducting qubits have demonstrated a stronger performance compared to other modalities. As a result, we opted to conduct a direct comparison of their theoretical performance with that of a custom naive minimum-finding function.
\begin{figure}[ht]
    \centering
    \includegraphics[width=8cm]{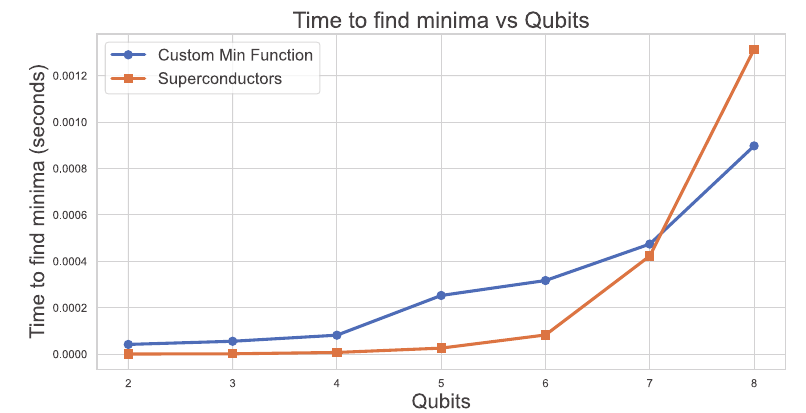}
    \caption{Minima finding performance theoretical performance of superconducting quantum circuit vs classical custom naive minima finding a function on list lengths of 4 to 256 ($2^2$ to $2^8$)}
    \label{fig:superconductor_comp}
\end{figure}

It can be observed from Figure \ref{fig:superconductor_comp} that as the list size increases, the total cost for superconductors and the custom naive minima function both grow. However, when analyzing these values closely, we can identify a potential for superconducting qubits to match the performance of the classical minima finding function.

As the list size expands, the total cost for superconducting qubits matches the performance of the classical minima function. This suggests that, in terms of gate time performance, superconducting qubits have the potential to offer a viable alternative to the classical method. While there are still differences in the execution times between the two approaches, the data indicates that the gap between them is competitive, particularly on the smaller lists providing a promising insight into the potential capabilities of superconducting qubits.

Given these observations, future research should investigate the possibility of developing hardware support for the minima-finding circuit using superconducting qubits. By accelerating the performance of this operation, researchers could unlock new levels of efficiency and optimization in quantum computing applications. In addition, the exploration of hardware-based solutions for minima-finding could pave the way for further advancements in quantum algorithms and problem-solving techniques. This line of inquiry holds great potential for enhancing the capabilities of quantum computing and pushing the boundaries of what is possible with this emerging technology.

\subsection{Performance of Quantum-enhanced Dijkstra} \label{ss:perf_q_dijkstra}
We tested our algorithm on 9 routes, which are divided into three groups: Short Domestic Flights, Long Domestic Flights, and Long International Flights. The routes are generated through our pipeline, with the addition of 2 midpoints and 5 perturbed nodes. The final graph is presented in the Nodes and Edges column of Table 
\begin{table}[h]
\centering
\caption{Domestic and International Routes}
\begin{tabular}{|l|c|c|c|}
\hline
\multirow{2}{*}{\textbf{Route}} & \textbf{Distance} & \textbf{Waypoints/} & \textbf{Nodes/} \\
 & \textbf{(km)} & \textbf{Midpoints} & \textbf{Edges} \\
\hline
Philadelphia - Boston & 449.6 & 5/4 & 44/228 \\
New York - Chicago & 1188.5 & 16/7 & 146/840 \\
Los Angeles - Denver & 1385.6 & 15/11 & 128/732 \\
\hline
\textbf{Avg. Short Domestic} & 1007.90 & 12/7.33 & 106/600 \\
\hline
Chicago - Houston & 1490.4 & 17/12 & 164/948 \\
Washington D.C. - Las Vegas & 3354.6 & 27/20 & 272/1596 \\
New York - San Francisco & 4153.2 & 31/23 & 314/1848 \\
\hline
\textbf{Avg. Long Domestic} & 2999.40 & 25/18.33 & 250/1464 \\
\hline
New York - London & 5541.1 & 29/14 & 248/1452 \\
Paris - Beijing & 8191.3 & 40/32 & 674/4008 \\
Mumbai - Sydney & 10162.1 & 83/31 & 422/2496 \\
\hline
\textbf{Avg. International} & 7964.83 & 50.67/25.67 & 448/2652 \\
\hline
\end{tabular}
\label{tab:routes}
\end{table}

We analyze three quantum computing technologies - Neutral Atoms, Ion Traps, and Superconductors - by comparing their gate operation times (Table \ref{tab:gate-times}). The data includes average times in milliseconds for four types of gates - $X, H, CNOT$, and $Z$ measurement - across three categories of flights for each technology

Superconductors exhibit the fastest gate operation times, making them advantageous for executing quantum algorithms requiring fast gate operations to minimize error accumulation. Superconducting qubits can be fabricated with semiconductor manufacturing techniques, enabling easier integration with existing infrastructure. However, they have shorter coherence times and can be sensitive to various types of noise, leading to the need for more advanced error correction techniques \cite{preskill2018quantum}.

Our analysis shows that Superconductors offer a significant time advantage, as seen on Table \ref{tab:gate-times}, over Neutral Atoms and Ion Traps for these gate operations. However, the advantages of Neutral Atoms and Ion Traps, such as scalability, long coherence times, and high-fidelity operations, cannot be overlooked. They may be more suitable for certain quantum computing applications.

\begin{table}[h]
\centering
\caption{Gate Times for Quantum Technologies\cite{suchara2013comparing}}
\begin{tabular}{lcccc}
\toprule
 & \multicolumn{4}{c}{Gate Times (milliseconds)} \\
\cmidrule(lr){2-5}
Technology & X & H & CNOT & Z meas. \\
\midrule
\multicolumn{5}{c}{Average Short Domestic} \\
\midrule
Superconductors & 0.571 & 0.0924 & 0.3056 & 0.0440 \\
Neutral Atoms & 152.3 & 46.06 & 157.9 & 352.0 \\
Ion Traps & 285.6 & 92.40 & 1667.0 & 440.0 \\
\midrule
\multicolumn{5}{c}{Average Long Domestic} \\
\midrule
Superconductors & 0.6314 & 0.0988 & 0.3439 & 0.0461 \\
Neutral Atoms  & 168.4 & 49.23 & 177.7 & 368.8 \\
Ion Traps & 315.7 & 98.76 & 1876.0 & 461.0 \\
\midrule
\multicolumn{5}{c}{Average International} \\
\midrule
Superconductors & 0.6355 & 0.0996 & 0.3065 & 0.0463 \\
Neutral Atoms & 169.5 & 49.65 & 158.4 & 370.4 \\
Ion Traps & 317.7 & 99.60 & 1672.0 & 463.0 \\
\bottomrule
\end{tabular}
\label{tab:gate-times}
\end{table}

\subsection{Performance enhancing techniques}
The results discussed in the preceding subsections show the potential of quantum-enhanced Dijkstra's algorithm in optimizing flight paths. To fully harness the capabilities of quantum computing for flight path optimization problems, or any other applications, additional innovations must be developed. In the following sections, we will address some of these challenges and explore potential resolutions.
\subsubsection{Hardware results and error mitigation}
\begin{figure}[!htb]
    \centering
    \includegraphics[width=8cm]{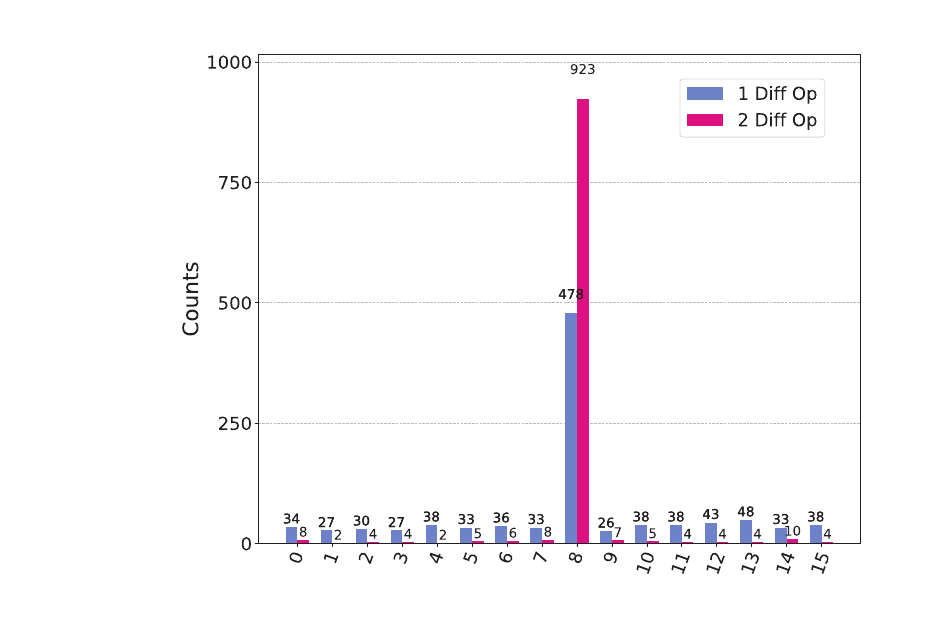}
    \caption{IBM QASM Simulator}
    \label{fig:qasmsim}
\end{figure}

\begin{figure}[!htb]
    \centering
    \includegraphics[width=8cm]{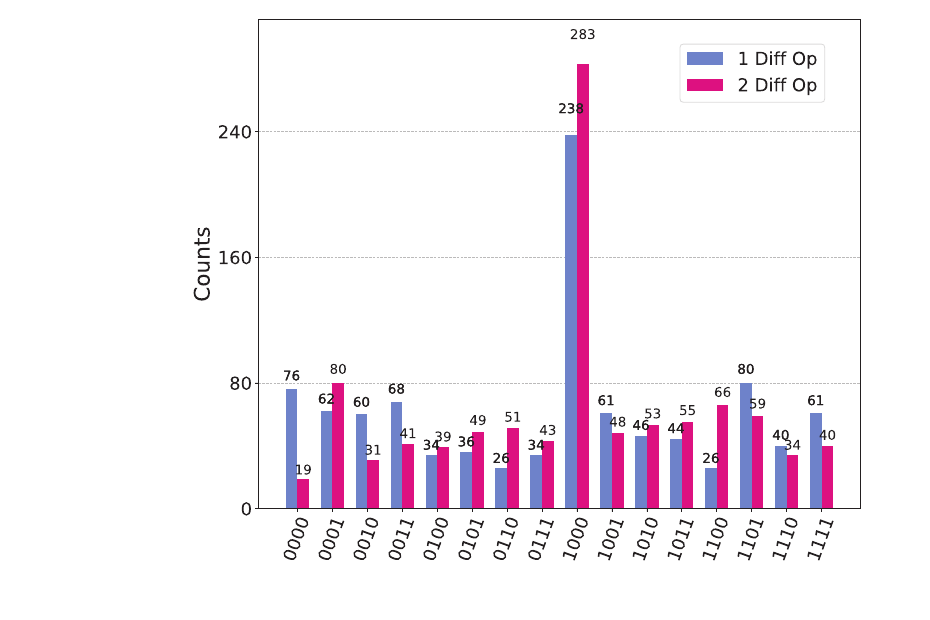}
    \caption{IBM Quito}
    \label{fig:hardwareRun}
\end{figure}

\begin{figure}[!htb]
    \centering
    \includegraphics[width=8cm]{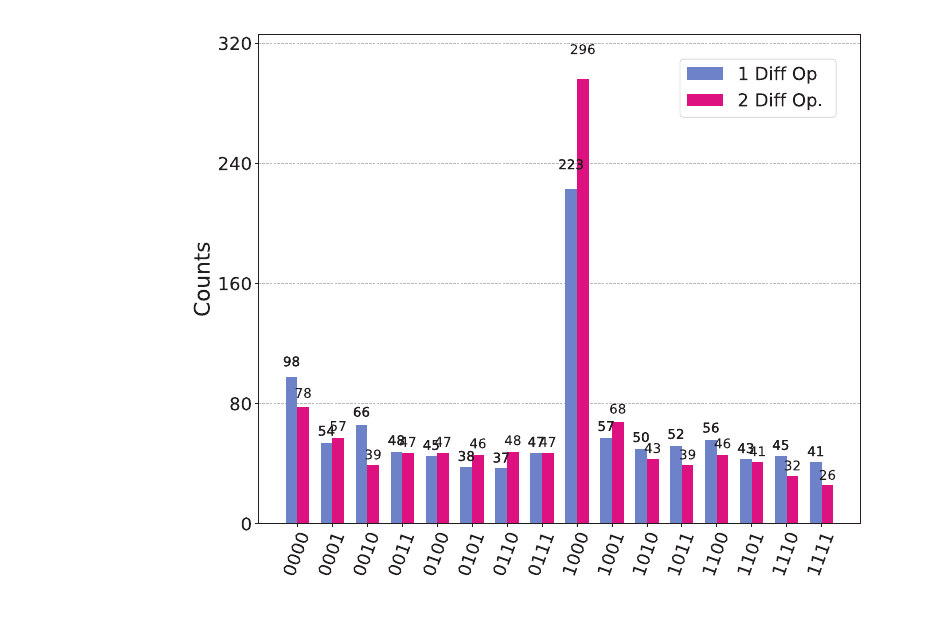}
    \caption{Fire Opal}    
    \label{fig:hardwareRun2}
\end{figure}

Figure \ref{fig:hardwareRun} displays the outcomes of executing Grover's algorithm for four qubits using a simulator, IBM's Quito machine, and IBM's Quito machine with Q-CTRL's Fire Opal error correction software. Two runs were conducted for each setup, incorporating one and two diffusion operators. The fidelity of the results obtained from the actual quantum computer (Figure \ref{fig:hardwareRun},\ref{fig:hardwareRun2}) is considerably lower compared to those derived from the simulator (Figure \ref{fig:qasmsim})\cite{qctrlblog2021}. Enhancing the hardware outcomes is feasible through the utilization of error mitigation software like Fire Opal; however, this introduces overhead, as circuits necessitate recompilation. Disregarding the network costs associated with the current architecture, generating new pulse sequences for a given quantum circuit remains essential. We anticipate that the future will bring a more seamless integration of error mitigation software, which should address both issues.

It is important to acknowledge that as quantum hardware matures, the number of shots required for Grover's algorithm will decrease. The shot count exhibits an interesting interplay with the number of diffusion operator iterations and the hardware quality. As evident from the four-qubit simulator results, incorporating an additional diffusion operator enhances the probability of obtaining the correct string. While extra diffusion operators could further reduce the shots, their impact on real quantum hardware is minimal. This diminished improvement results from noise, as the introduction of extra diffusion operators leads to lengthier circuits, allowing noise to offset the enhancement. As quantum hardware and error mitigation software continues to progress, determining the optimal combination of diffusion operators and shots to minimize overall execution time will be crucial.

\subsubsection{Pre-compiled amplitude amplification}
As demonstrated in section \ref{ss:perf_q_dijkstra}, superconducting architecture, characterized by fast gate execution times, appears to be the most promising modality for Grover's algorithm. In addition to gate execution times, significant time expenses in current architectures are attributed to network costs and the construction of quantum circuits in Python. For example, assembling four-qubit circuits for Grover's algorithm requires 124 ms. Although on-premise hardware could substantially reduce network costs, further innovations are needed to streamline circuit building. One possible approach involves using pre-compiled circuits for amplitude amplification circuits. It should be feasible to develop pre-compiled circuits for QRAM \cite{giovannetti2008quantum} architecture.
\subsubsection{Parallelizing the quantum circuits}
The execution time of the amplitude amplification algorithm could potentially be reduced by parallelizing circuits on more expansive quantum devices. IBM has unveiled its 433-qubit machine in 2022 \cite{ibmnewsroom2022}. It is plausible that such a device could concurrently execute at least 30 four-qubit circuits. This suggests that for a four-qubit circuit, data corresponding to 30 shots could be acquired in a single clock cycle. Parallelization may emerge as a vital technique, given that measurement and qubit reset processes are time-intensive.

\section{Conclusion} \label{sec:conc}
In this study, we investigated the application of quantum algorithms to accelerate solutions to the flight path optimization problem. We introduced an implementation of quantum-enhanced Dijkstra and benchmarked it on 3 different modalities. We determined that the superconducting qubit architecture is the most suitable for our implementation. Our analysis suggests that quantum computing holds great promise for the future, though additional innovation may be necessary for practical applications.

\section*{Acknowledgements} We acknowledge insightful discussions with Scott Campbell, Kenneth Heitritter, Robert Keele, Jasper Simon Krauser, Zsolt Lattman, and Ricky Young.

\bibliographystyle{ieeetr}
\bibliography{main}

\end{document}